\documentclass[sigconf]{acmart}

\usepackage{amsmath,amsfonts}
\usepackage{graphicx}
\usepackage{textcomp}
\usepackage{xcolor}
\usepackage{natbib}
\usepackage{hyperref}
\usepackage{multirow}
\usepackage{booktabs}
\usepackage{float}
\usepackage{mathrsfs}
\usepackage{subfigure}
\usepackage{lineno}
\usepackage{algorithm}
\usepackage{algorithmic}
\usepackage{color}

\newtheorem{theorem}{Theorem}[section]

\newtheorem{prop}{Proposition}[section]

\newtheorem{exmp}{Example}[section]
\newcounter{hypA}
\newenvironment{hypA}{\refstepcounter{hypA}\begin{itemize}
  \item[({\bf A\arabic{hypA}})]}{\end{itemize}}
\newcounter{hypB}

\newcounter{hypC}

\usepackage{color-edits}
\definecolor{redorange}{RGB}{255, 68, 51}
\addauthor[Denglu]{dl}{redorange}

\AtBeginDocument{%
  \providecommand\BibTeX{{%
    \normalfont B\kern-0.5em{\scshape i\kern-0.25em b}\kern-0.8em\TeX}}}

\setcopyright{acmcopyright}
\copyrightyear{2024}
\acmYear{2024}
\acmDOI{XXXXXXX.XXXXXXX}

\acmConference[KDD'24]{the ACM SIGKDD Conference on Knowledge Discovery and Data Mining 2024}{August 25--29, 2024}{Barcelona, Spain}

\acmPrice{15.00}
\acmISBN{978-1-4503-XXXX-X/18/06}

\acmSubmissionID{836}

\begin{document}

\title{
Unbiased Estimation for Total Treatment Effect Under Interference Using Aggregated Dyadic Data
}


\author{Lu Deng}
\affiliation{%
  \institution{Tencent, Inc.}
  \city{Shenzhen, Guangdong}
  \country{China}}
\email{adamdeng@tencent.com}

\author{Yilin Li}
\affiliation{%
  \institution{Peking University}
  \city{Beijing}
  \country{China}}
\email{yilinli@pku.edu.cn}

\author{JingJing Zhang}
\affiliation{%
  \institution{Tencent, Inc.}
  \city{Shenzhen, Guangdong}
  \country{China}}
\email{broccozhang@tencent.com}

\author{Yong Wang}
\affiliation{%
  \institution{Tencent, Inc.}
  \city{Shenzhen, Guangdong}
  \country{China}}
\email{darwinwang@tencent.com}

\author{Chuan Chen}
\affiliation{%
  \institution{Tencent, Inc.}
  \city{Shenzhen, Guangdong}
  \country{China}}
\email{chuanchen@tencent.com}

\renewcommand{\shortauthors}{Lu Deng, Yilin Li, Jingjing Zhang, Yong Wang, and Chuan Chen}

\begin{abstract}
In social media platforms, user behavior is often influenced by interactions with other users, complicating the accurate estimation of causal effects in traditional A/B experiments. This study investigates situations where an individual's outcome can be broken down into the sum of multiple pairwise outcomes, a reflection of user interactions. These outcomes, referred to as dyadic data, are prevalent in many social network contexts. Utilizing a Bernoulli randomized design, we introduce a novel unbiased estimator for the total treatment effect (TTE), which quantifies the difference in population mean when all individuals are assigned to treatment versus control groups. We further explore the bias of our estimator in scenarios where it is impractical to include all individuals in the experiment, a common constraint in online control experiments. Our numerical results reveal that our proposed estimator consistently outperforms some commonly used estimators, underscoring its potential for more precise causal effect estimation in social media environments.
\end{abstract}





\maketitle

\section{Introduction}
Online control experiments provide robust methodologies for tech companies to assess the impact of innovative product strategies \cite{kohavi1, kohavi2}. In such experiments, the population is randomly divided into a treatment group, which receives the treatment under investigation, and a control group, which maintains the status quo. The treatment effect is then estimated by comparing the difference in outcomes between these two groups post-treatment. Classical causal inference techniques for estimating these effects rely on the Stable Unit Treatment Value Assumption (SUTVA). SUTVA posits that an individual's outcome is independent of the treatment assignments of others, ensuring that there is no interference between units. In scenarios where SUTVA is valid, the difference-in-means estimator or Horvitz-Thompson estimator  provides an unbiased estimate of the treatment effect \cite{imbens}.

However, social network settings commonly violate the SUTVA, as an individual's behavior can be influenced by their social connections \cite{eckles, abadie2019, kakhki, gui, ugander}. This violation may introduce bias into estimates from traditional randomized experiments. For instance, consider a social media platform testing a feature to promote user messaging. This feature could inadvertently stimulate friends in the control group to increase their messaging activity. Consequently, both the treatment and control groups exhibit increased activity, potentially leading to an underestimation of the treatment's actual effect. This phenomenon, known as network effects, is also referred to as interference, spillover, or peer influence effects. Our primary focus is to estimate the difference in population mean outcomes when all individuals are assigned to either the treatment or control group.

Causal inference amidst interference has attracted increasing interest from academia and industry in recent decades. There are two primary methods to address this: the design-based approach and the model-assisted approach. The design-based approach aims to reduce bias in effect estimation by strategically assigning treatments. One such method is graph cluster randomization \cite{gui, ugander, Ugander2023, hayes}, which groups a network into clusters and treats each cluster as a whole. While this can reduce bias, it may also decrease the precision of results and is not suitable for highly interconnected networks. Another method is LinkedIn's ego-cluster experiment \cite{saintjacques}, which focuses on individuals and their direct connections. However, this method has low power and may still introduce bias. The model-assisted approach involves constructing mathematical models of potential outcomes and estimating parameters to understand effects \cite{ sarndal2003, basse2017, toulis2013, basse2015optimal, sussman2017elements, forastiere2021identification, saint2019method}. The exposure model \cite{manski2013, aronow2012} is an example of this approach, simplifying the treatment information into "exposures" to study causal effects. However, this approach assumes that these exposures fully encapsulate the causal relationships, which may not always be the case. Recent research has explored the implications of incorrect models and found that some estimators can still be reliable under broader interference conditions, but determining the conditions remains challenging \cite{chin2018, savje2021average, savje2023causal, leung2022causal}.

This work's primary contribution is the development of novel estimators for interference within the context of dyadic data. Dyadic data, capturing interactions between pairs of users (dyads), enables the investigation of complex dynamics and relationships among users. For example, in social network analysis, dyads may refer to individuals within a network. Utilizing dyadic information allows researchers to gain insights into intricate interactive behavior. Dyadic data analysis is crucial across various scientific disciplines, including economics \cite{portes2005}, social sciences \cite{apicella2012}, and biomedical sciences \cite{luke2007}. 
To our knowledge, we are the first to propose using aggregated dyadic data to estimate treatment effects amidst interference. We introduce and analyze several new estimators that incorporate dyadic structures under different experimental scenarios. The proposed framework is easy to set up and is widely run at Weixin's experiment platform. Our simulation results validate the theorems and demonstrate the performance of the proposed estimators. In summary, our contributions include:
\begin{enumerate}
\item Formalizing the potential outcome model with dyadic outcomes and providing an unbiased estimator of the total treatment effect under a simple randomized experiment.
\item Offering a detailed analysis of our estimator when the experiment includes only a small fraction of the population.
\item Proposing a novel experimental scheme that uses clustering to reduce the bias of the estimator in subpopulation experiments.
\item Conducting systematic simulations on two social networks and comparing our method with traditional estimators, thereby providing a credible reference for our method's effectiveness.
\end{enumerate}

The structure of this paper is as follows: Section \ref{sec2} provides an initial discussion on the potential outcome model and network interference. Section \ref{sec3} details our proposed estimators. Section \ref{sec4} presents simulation studies and a real-world example that corroborates our theoretical findings. Section \ref{sec5} concludes with a thorough discussion. Appendix A contains the proofs of some of our theoretical results.

\section{Problem Setting}\label{sec2}
In this section, we'll formalize the problem of causal inference under interference, discuss some commonly used assumptions, and describe the structure of dyadic data.

\subsection{Potential outcome model}
Consider a finite population of $n$ individuals. Let $W_i \in \{0,1\}$ represent the random binary treatment assignment for individuals $i$, where 1 and 0 denote treatment and control, respectively. Let $\mathbf{W} = (W_1,\dots,W_n)^{\intercal} \in \Omega \overset{\Delta}{=}  \{ 0, 1 \}^n$, a vector of length $n$, be the experiment assignment vector for all individuals. Let $Y_i(\mathbf{W})$ denote the potential outcome for individual $i$ if the treatment vector $\mathbf{W}$ is implemented. Our goal is to estimate the total treatment effect (TTE). TTE is defined as the difference in population mean when all individuals are placed in treatment versus control, which can be expressed as 
\begin{equation}\label{equa: tte}
\tau \overset{\Delta}{=} \frac{1}{n} \sum_{i=1}^n \Big[Y_i(\mathbf{W} = \mathbf{1}) - Y_i(\mathbf{W} = \mathbf{0})\Big].
\end{equation}
where $\mathbf{1} = (1,\dots,1)^{\intercal}, \mathbf{0} = (0,\dots,0)^{\intercal}$. In what follows, expectations $\mathbb{E}$ are over the treatment assignment only, potential outcomes are considered fixed. Under SUTVA, $Y_i(\mathbf{W}) = Y_i(W_i)$ for all $\mathbf{W} \in \Omega$. Therefore, TTE can be simplified as
\begin{equation}
    \tau = \frac{1}{n} \sum_{i=1}^n \Big[Y_i(W_i = 1) - Y_i(W_i = 0)\Big].    
\end{equation}
A commonly used estimator for $\tau$ is the difference-in-means estimator, defined by
\begin{equation}
\hat{\tau}_{dim} = \frac{\sum_{i=1}^n W_i Y_i}{\sum_{i=1}^n W_i}  - \frac{\sum_{i=1}^n (1-W_i) Y_i}{\sum_{i=1}^n (1-W_i)}.
\end{equation}
Under Bernoulli randomization or complete randomization, $\hat{\tau}_{dim}$ is an unbiased estimator for $\tau$ \cite{imbens}. However, when interference exits, $\hat{\tau}_{dim}$ is biased and fails to provide a valid estimate of  $\tau$\cite{eckles, gui}.

\subsection{Network Graph}
In the following sections, we consider that the interference can be characterized by a directed graph $G = (V, E)$, where the node set $V = \{1, 2, \dots, n\}$ represents the individuals, and $E=\{ (i, j)\}$ is the collection of edges that represent interference between individuals. We use $|E|$ to denote the cardinality of $E$, which is the number of edges in the graph. Each graph is represented by an adjacency matrix. Let $A$ be an $n\times n$ matrix with entries $(A_{ij})_{1\leq i,j \leq n}$ where $A_{ij} \in \{0, 1\}$ with $A_{ii} = 0$, and for $i\neq j$,
\begin{equation}
    A_{ij} \overset{\Delta}{=} \begin{cases}
    0 & \text{no edge between $i$ and $j$} \\
    1 & \text{direct edge from $i$ to $j$}.
    \end{cases}    
\end{equation}
We let $N_u(j) \overset{\Delta}{=} \{i: A_{ij} = 1 \}$ and $N_d(j) \overset{\Delta}{=} \{i: A_{ji} = 1 \}$ represent the upstream neighbors and downstream neighbors set of user $j$ with respect to the graph, respectively. Throughout the article, we assume that the network graph $G$ is fixed and known.

\subsection{Neighborhood Interference}
The neighborhood interference assumption \cite{Ugander2023, liu2022adaptive, yuan2022two, forastiere2022estimating, cortez2022staggered} posits that an individual's outcome depends solely on the treatment assignments of their direct neighbors within a specified network. This assumption is frequently utilized when addressing interference, as it simplifies the problem by focusing on local interactions.
\begin{hypA}
\label{hyp:neighbor_interference}
(Neighborhood Interference) The potential outcomes only depend on one's own and neighbors' treatment assignments: $Y_i (\mathbf{W}) = Y_i (\mathbf{W}^{\prime})$ if $W_i = W_i^{\prime}$ and $W_j = W_j^{\prime}$ for all $j \in N_u(i) \cup N_d(i)$.
\end{hypA}

Assuming neighborhood interference, the Horvitz-Thompson estimator combined with exposure mapping can be employed to estimate the total treatment effect \cite{aronow2017estimating}. However, the practical application of such estimators is often hindered by their substantial variances or their dependence on stringent exposure mapping assumptions.

\subsection{Dyadic data setup}
To estimate the total treatment effect, we propose an assumption that is applicable in real-world scenarios and pertains to potential outcomes. We denote the network graph characterizing the interference as $G = (V, E)$, and adopt the neighborhood interference assumption (A\ref{hyp:neighbor_interference}). In a given scenario, each direct edge $(i,j) \in E$ is associated with an outcome, represented as $z_{i,j}$. The outcome for unit $j$, denoted as $Y_j$, is defined as the summation of these associated outcomes from its upstream neighbors. 
\begin{equation}
    Y_j = \sum_{i \in N_u(j)} z_{i,j}.
\end{equation}

\begin{exmp}\label{exmp:pair}
Within the framework of content sharing and forwarding, consider a scenario where a user $i$ forwards a video to another user $j$. Upon receiving the video, user $j$ clicks on it, leading to the sharing page, and then watches the video. This action results in a stay duration, denoted by $z_{i,j}$. The cumulative stay duration on the sharing page for user $j$ over a specific period is calculated by summing up the stay durations $z_{i,j}$, each of which is generated by clicking on videos forwarded by friends. In the absence of any forwarding or clicking interaction between user $i$ and user $j$, we assign $z_{i,j}$ a value of 0, signifying that no stay duration is generated under such circumstances.
\end{exmp}

\begin{figure}[htbp]
\centering
\includegraphics[width=0.45\textwidth]{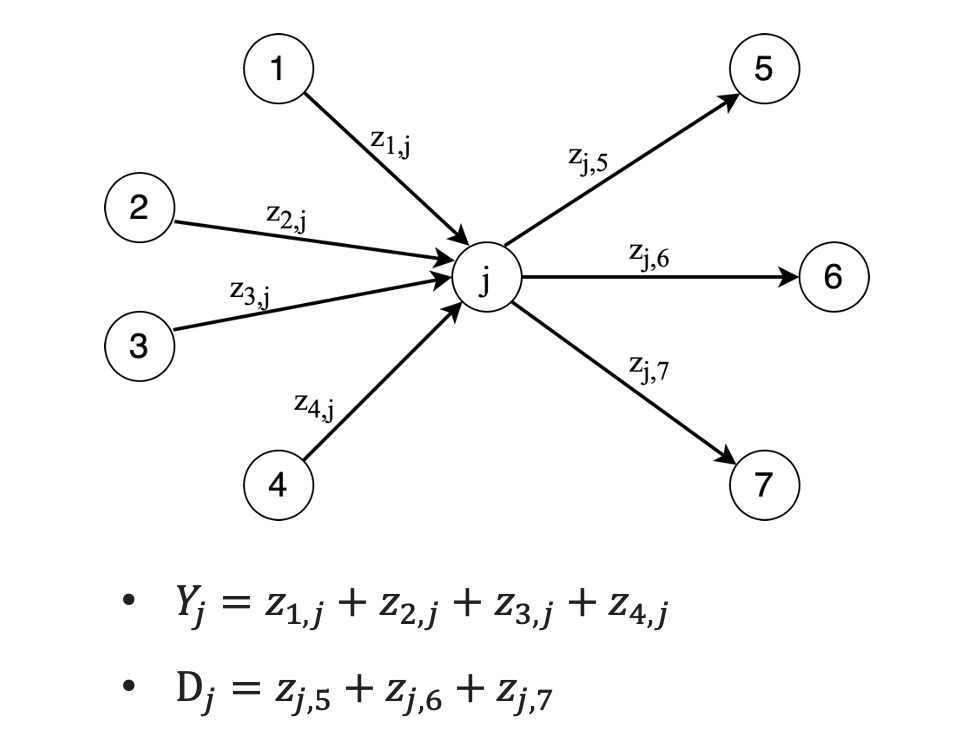}
\caption{An illustration of the calculation of $Y_i$ and $D_i$. Suppose node $i$ has 7 neighbor nodes, where nodes 1,2,3,4 are upstream neighbors, and nodes 5,6,7 are downstream neighbors. Each direct edge is associated with an outcome. $Y_j$ is the summation of the paired outcomes from upstream neighbors, which is $z_{1,j} + z_{2,j} + z_{3,j} + z_{4,j}$. $D_j$ is the summation of the paired outcomes from downstream neighbors, which is $z_{j,5} + z_{j,6} + z_{j,7}$.}
\label{fig:diffusion_metric}
\end{figure}

It's important to note that $z_{i,j}$ does not necessarily need to be symmetric for the pair $(i,j)$, which means that $z_{i,j} \neq z_{j,i}$ is permissible. Additionally, we set $z_{i,i} = 0$, indicating that we are considering a graph without a self-loop. The value of $z_{i,i}$ will influence the properties of our estimators, a point that will be illustrated in the subsequent section. For each unit $j$, we denote $D_j$ as the summation of paired outcomes from downstream neighbors:
\begin{equation}
    D_j = \sum_{i \in N_d(j)} z_{j,i}.
\end{equation}
We refer to $D_j$ as the diffusion metric, as it characterizes the spillover effect from individual $j$ diffusing to all of his neighbors. Moreover, we refer to $Y_j$ and $D_j$ as aggregated dyadic data. For a more detailed understanding of how $Y_i$ and $D_i$ are calculated, please refer to Figure \ref{fig:diffusion_metric}, which provides a visual representation of this process.

The forms of $Y_i$ and $D_i$ exhibit a kind of duality: one is a summation from upstream neighbors, and the other is from downstream neighbors. In fact, it can be shown that $\sum_{j=1}^n Y_j = \sum_{j=1}^n D_j$. Thus, the total treatment effect defined by $Y_i$ and $D_i$ are identical. If we denote $z_{i,j}(\mathbf{W}), Y_j(\mathbf{W})$ and $D_j(\mathbf{W})$ as the counterfactual outcomes under treatment vector $\mathbf{W}$, then
\begin{equation*}
\tau = \frac{1}{n} \sum_{i=1}^n \Big[Y_i(\mathbf{W} = \mathbf{1}) - Y_i(\mathbf{W} = \mathbf{0})\Big] = \frac{1}{n} \sum_{i=1}^n \Big[D_i(\mathbf{W} = \mathbf{1}) - D_i(\mathbf{W} = \mathbf{0})\Big].
\end{equation*}
In past studies, researchers have only used $Y_i$ to estimate the total treatment effect. Our work demonstrates that we can leverage the information provided by $D_j$ to construct a more accurate estimate.

\section{Methodology}\label{sec3}
In this section, we will delineate the fundamental assumption that forms the basis of our approach, introduce the novel estimator we have developed, and conduct an in-depth analysis of the performance of this estimator under a variety of conditions.

\subsection{Dyadic Interference}
Our overarching framework necessitates an interference assumption on the dyadic outcome $z_{i,j}(\mathbf{W})$.
\begin{hypA}
\label{hyp:pair}
For each $(i,j)$, the potential outcomes $z_{i,j}(\mathbf{W})$ only depend on unit $i$ and unit $j$'s treatment assignments: for any $\mathbf{W}, \mathbf{W}^{\prime} \in \Omega$, if $W_i = W_i^{\prime}$ and $W_j = W_j^{\prime}$, then $z_{i,j} (\mathbf{W}) = z_{i,j} (\mathbf{W}^{\prime})$.
\end{hypA}
Assumption (A\ref{hyp:pair}) embodies a specific form of the neighbor interference assumption, which is plausible considering that most dyadic data originate from pairwise interactions between units. Under this assumption, we further propose that $z_{i,j}(\mathbf{W})$ can be modeled as follows:
\begin{equation}\label{equa: zijmodel}
    z_{i,j}(\mathbf{W}) =  \alpha_{i,j} + \beta_{i, j} W_i + \gamma_{i, j} W_j + \zeta_{i,j} W_i W_j.
\end{equation}
The parameters $\beta_{i,j}$ and $\zeta_{i,j}$ are intended to capture network interference, specifically addressing the potential spillover effect. On the other hand, $\gamma_{i,j}$ is designed to capture the direct effect of treatment. 

Given \eqref{equa: zijmodel}, we can expand $Y_j(\mathbf{W})$ and $D_j(\mathbf{W})$ as
\begin{equation}\label{equa: yjmodel}
Y_j(\mathbf{W}) = \sum_{i \in N_u(j)}  \Big( \alpha_{i,j} + \beta_{i,j} W_i + \gamma_{i,j} W_j + \zeta_{i,j} W_i W_j \Big),
\end{equation}
and 
\begin{equation}\label{equa: djmodel}
D_j(\mathbf{W}) = \sum_{i \in N_d(j)}  \Big( \alpha_{j,i} + \beta_{j,i} W_j + \gamma_{j,i} W_i + \zeta_{j,i} W_i W_j \Big).
\end{equation}
The total treatment effect estimand $\tau$ takes the value
\begin{align*}\
\tau = & \,\,  \frac{1}{n} \sum_{i=1}^n \Big[Y_i(\mathbf{W} = \mathbf{1}) - Y_i(\mathbf{W} = \mathbf{0})\Big]   \\
= & \,\, \frac{1}{n}\sum_{j=1}^n \sum_{i\in N_u(j)} \Big(\beta_{i,j} + \gamma_{i,j} + \zeta_{i,j}\Big).    
\end{align*}
Estimating these parameters presents a significant challenge due to the substantial number of model parameters, which scales with the number of edges in the network $O(|E|)$. Therefore, we do not estimate these parameters in practice. 

We can rewrite $\eqref{equa: yjmodel}$ as follows:
\begin{equation*}
Y_j(\mathbf{W}) = \alpha_j + \gamma_j  W_j + \sum_{i \in N_u(j)} \Big(\beta_{i,j} W_i +  \zeta_{i,j} W_i W_j\Big),
\end{equation*}
where $\alpha_j = \sum_{i \in N_u(j)} \alpha_{i,j}$ and $\gamma_j = \sum_{i \in N_u(j)} \gamma_{i,j}$. By setting the parameter $\zeta_{i,j}=0$, the model simplifies to the heterogeneous additive network effects model, as discussed by Yu et al.\cite{yu2022estimating} in their 2022 paper. This demonstrates that our model is a more general form capable of encompassing the model proposed by Yu and colleagues when certain conditions are applied.

To simplify the formula, we assume that $z_{i,j} = 0 $ if $(i,j) \notin E$. Then the TTE $\tau$ can be expressed as follows:
\begin{equation}\label{equa: tteinmodel}
\tau =  \frac{1}{n}\sum_{i\neq j}\Big(\beta_{i,j} + \gamma_{i,j} + \zeta_{i,j}\Big).
\end{equation}

\subsection{Estimator}\label{subsec:estimator}
Consider a simple Bernoulli randomization experiment, where each unit in the experiment is randomly assigned to either the treatment group or the control group. Let $\pi = \mathbb{P}(W_i = 1)$ represent the treatment probability. We define two Horvitz-Thompson (HT) estimators:
\begin{equation}\label{equa: htate}
\hat{\tau}^{1}(\pi) = \frac{1}{n} \sum_{i=1}^n\frac{W_i Y_i}{\pi} - \frac{1}{n} \sum_{i=1}^n \frac{(1-W_i)Y_i}{1-\pi},
\end{equation}
and
\begin{equation}\label{equa: htdiffuse}
\hat{\tau}^{2}(\pi) = \frac{1}{n} \sum_{i=1}^n\frac{W_i D_i}{\pi} - \frac{1}{n} \sum_{i=1}^n \frac{(1-W_i)D_i}{1-\pi}.
\end{equation}
The parameter $\pi$ in the estimators suggests that the properties of the estimators are dependent on $\pi$. The $\hat{\tau}^{1}(\pi)$ is a commonly used estimator in causal inference. It is widely recognized that $\hat{\tau}^{1}(\pi)$ is biased for the total treatment effect $\tau$ when interference exists \cite{eckles, gui}.

Finally, we define a new estimator $\hat{\tau}(\pi)$ as the sum of $\hat{\tau}^1(\pi)$ and $\hat{\tau}^1(\pi)$. That is,
\begin{equation}\label{equa: htadd}
\hat{\tau}(\pi) \overset{\triangle}{=} \hat{\tau}^1(\pi) + \hat{\tau}^2(\pi).
\end{equation}
We use $\hat{\tau}(\pi)$ to estimate the total treatment effect $\tau$. We will next demonstrate the properties of these three estimators under various scenarios. All the proofs can be found in the appendix.

\subsection{Full population experiment}
We begin by considering the simplest scenario in which the entire population participates in the experiment. The following result is provided: 
\begin{prop}\label{prop: prop1}
$\mathbb{E}\big[\hat{\tau}^1(\pi)\big] = \frac{1}{n} \sum_{i \neq j} \big( \gamma_{i,j} + \zeta_{i,j}\pi \big).$
\end{prop}
The estimator's bias is given by $ - \frac{1}{n} \sum_{i \neq j} \Big( \beta_{i,j} + \zeta_{i,j}(1-\pi) \Big)$, which can scale with the average spillover effect across the edges between the treated and control groups. In practice, the sign of $\beta_{i,j}$ and $\zeta_{i,j}$ may not be identical, implying that we do not even know whether our estimate is an overestimate or an underestimate. Thankfully, with the assistance of $\hat{\tau}^{2(\pi)}$, we can unbiasedly estimate the total treatment effect. This additional information can then be employed to significantly simplify the estimation of causal effects. The following result characterize the bias of $\hat{\tau}^{2}(\pi)$.
\begin{prop}\label{prop: prop2}
$\mathbb{E}\big[\hat{\tau}^2(\pi)\big] = \frac{1}{n} \sum_{i \neq j} \big(\beta_{i, j} + \zeta_{i,j} \pi\big) .$
\end{prop}
$\hat{\tau}^{2}(\pi)$ is also a biased estimator for the total treatment effect. From Proposition \ref{prop: prop1} and \ref{prop: prop2}, we can infer that
$$\mathbb{E}\big[\hat{\tau}(\pi)\big] =  \frac{1}{n} \sum_{i \neq j} \big(  \beta_{i, j} + \gamma_{i, j} + 2\pi\zeta_{i,j}\big).$$
Consequently, for $\pi=0.5$, we derive $\mathbb{E}\big[\hat{\tau}(0.5)\big] = \frac{1}{n} \sum_{i \neq j} \big( \beta_{i, j} + \gamma_{i, j} + \zeta_{i,j} \big) = \tau$. Therefore, when $\pi=0.5$, $\hat{\tau}(0.5)$ serves as an unbiased estimator of the total treatment effect. We summarize the analysis above for the estimator $\hat{\tau}$ in the following theorem.
\begin{theorem}\label{thm: thm1}
Assume (A\ref{hyp:pair}) and that the experiment includes all units, the expectation of $\hat{\tau}(\pi)$ is
$$\mathbb{E}\big[\hat{\tau}(\pi)\big] =  \frac{1}{n} \sum_{i \neq j} \big( \beta_{i, j} + \gamma_{i, j} + 2\pi\zeta_{i,j}\big).
$$
When $\pi=0.5$, that is, when treatment and control assignments are equally probable, $\hat{\tau}(0.5)$ is an unbiased estimator of the total treatment effect $\tau$.
\end{theorem}

There are several important points to note in this subsection. Firstly, our proposed estimator relies heavily on assumption (A\ref{hyp:pair}), which states that $z_{i,j}$ depends solely on the treatment assignments of unit $i$ and unit $j$. This assumption is akin to the neighborhood interference assumption. While a similar approach to the method in \cite{athey} can be used to test assumption A\ref{hyp:pair}, it may not be as efficient. Secondly, we assume that  $z_{i,i} = 0$ for all $i$. This assumption is crucial for the unbiased property of $\hat{\tau}$. If we were to allow $z_{i,i} \neq 0$, then it can be shown that when $\pi = 0.5$, the bias of $\hat{\tau}$ is $\frac{1}{n}\sum_{i}(\beta_{i,i}+\gamma_{i,i}+\zeta_{i,i})$. However, even when $z_{i,i} \neq 0$, compared to $\hat{\tau}^1$, our proposed estimator $\hat{\tau}$ still shows an improvement in terms of bias, which further underscores its potential usefulness in practice. We leave the proof of this statement to the reader.

\subsection{Sub-population experiment}\label{subsec:subpopulation}
Theorem \ref{thm: thm1} provides an unbiased estimator for the total treatment effect. However, this holds only when we conduct an controlled experiment with the entire population. In reality, many industry experimental platforms run thousands of experiments simultaneously every day using a layered structure design (see \cite{tang2010, kohavi1, xiong2020} for more details about the layer and experiment structure). Typically, an experiment encompasses only a small fraction of users, often less than 10\% according to \cite{kohavi1}. This leads us to a natural question: Does the conclusion of Theorem \ref{thm: thm1} still holds when the experiment includes only a subset of the population? This is a critical question, as it directly impacts the applicability and relevance of our findings in real-world settings.

Let's suppose we randomly select a proportion $p(0 < p \leq 1)$ of the population to participate in the experiment. For each user $i$, we use $V_i \in \{0,1\}$ to denote whether user $i$ is included in the experiment or not. Users who are included in the experiment are then assigned to either the treatment or control group through a Bernoulli randomization process. Let's denote $\pi = \mathbb{P}(W_i = 1 | V_i = 1)$ as the probability of users in the experiment receiving the treatment. To simplify the problem, we'll assume that users who are not included in the experiment receive the control group strategy, i.e., $\mathbb{P}(W_i = 0 | V_i = 0)=1$. Under these assumptions, the two HT estimators become functions of $p$ and $\pi$:
\begin{equation}\label{equa: htate2}
\hat{\tau}^{1}(p, \pi) = \frac{1}{n} \sum_{i=1}^n\frac{V_iW_i Y_i}{p\pi} - \frac{1}{n} \sum_{i=1}^n \frac{V_i(1-W_i)Y_i}{p(1-\pi)},
\end{equation}
and
\begin{equation}\label{equa: htdiffuse2}
\hat{\tau}^{2}(p, \pi) = \frac{1}{n} \sum_{i=1}^n\frac{V_i W_i D_i}{p\pi} - \frac{1}{n} \sum_{i=1}^n \frac{V_i (1-W_i)D_i}{p(1-\pi)}.
\end{equation}
This setup allows us to explore how the proportion of the population included in the experiment and the probability of receiving treatment affect the bias of our estimators. The bias property of these two estimators are presented below.
\begin{prop}\label{prop: prop4}
$\mathbb{E}\big[\hat{\tau}^1(p, \pi)\big] = \frac{1}{n} \sum_{i \neq j} \big(\gamma_{j,i} + p\pi\zeta_{j,i} \big) .$
\end{prop}
\begin{prop}\label{prop: prop5}
$\mathbb{E}\big[\hat{\tau}^2(p, \pi)\big] = \frac{1}{n} \sum_{i \neq j} \big(\beta_{i,j}+ p\pi\zeta_{i,j}\big) .$
\end{prop}

\begin{figure*}[htbp]
\centering
\includegraphics[width=0.8\textwidth]{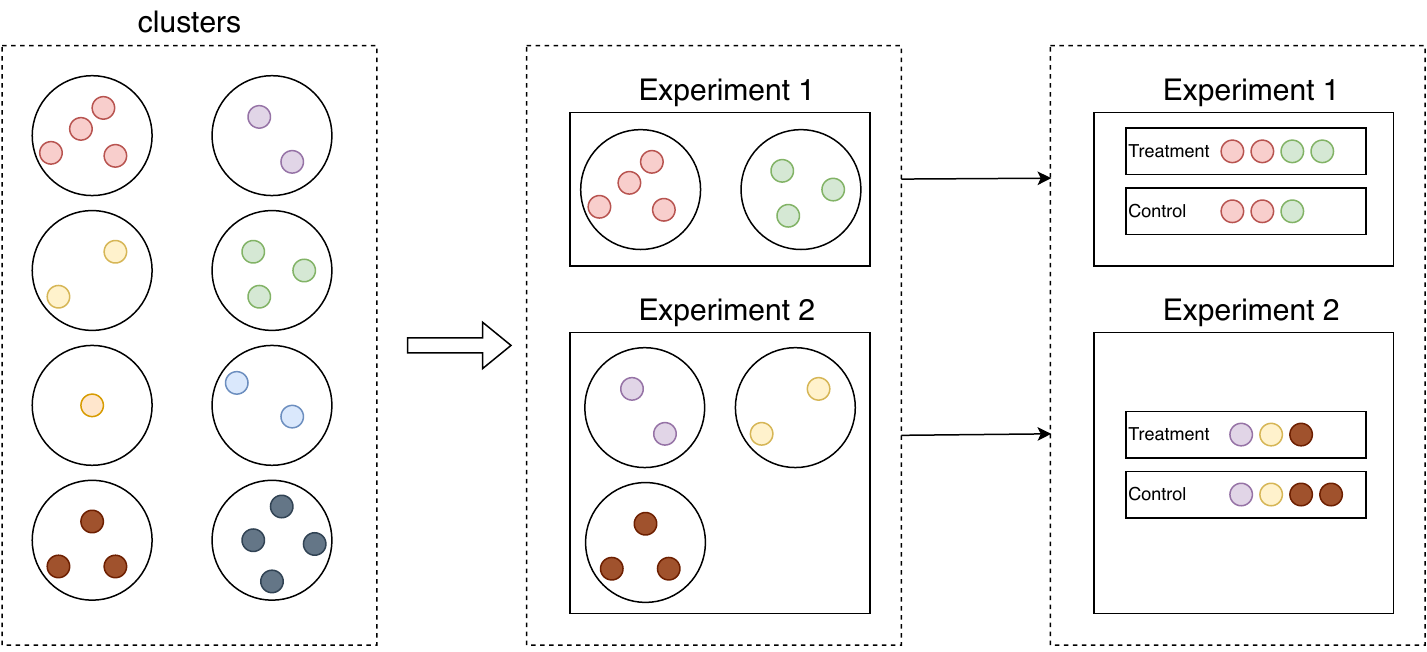}
\caption{Visualization of the two-stage experiment process.}
\label{fig:two_stage_expt}
\end{figure*}

Note that $\hat{\tau}(p, \pi) = \hat{\tau}^1(p, \pi) + \hat{\tau}^2(p, \pi)$, We have the following theorem about $\hat{\tau}(p, \pi)$.
\begin{theorem}\label{thm: thm2}
Assume (A\ref{hyp:pair}). The expectation of $\hat{\tau}(p, \pi)$ is
\begin{equation}
    \mathbb{E}\big[\hat{\tau}(p, \pi)\big] =  \frac{1}{n} \sum_{i \neq j} \big( \beta_{i, j} + \gamma_{i, j} + 2p\pi\zeta_{i,j}  \big).
\end{equation}
and the bias is 
\begin{equation}\label{equa:bias_sub}
    \mathbb{E}\big[\hat{\tau}(p, \pi) - \tau\big] = - \frac{1-2p\pi}{n} \sum_{i \neq j} \zeta_{i,j}.
\end{equation}
\end{theorem}

Theorem \ref{thm: thm1} can be seen as a special case of Theorem \ref{thm: thm2}. Specifically, when $p = 1$, meaning the entire population is included in the experiment, Theorem \ref{thm: thm2} simplifies to Theorem \ref{thm: thm1}. However, in general, we have $\sum_{i \neq j} \zeta_{i,j} \neq 0$. This implies that the estimator $\hat{\tau}(p, \pi)$ in a sub-population experiment (where $p<1$) is biased for the total treatment effect. Furthermore, there are $O(|E|)$ terms in the summands of $\sum_{i \neq j} \zeta_{i,j}$, which suggests that the bias is $O\Big(\frac{|E|}{n}\Big)$. When comparing the biases of the two estimators $\hat{\tau}(p, \pi)$ and $\hat{\tau}^{1}(p, \pi)$, we find that the difference in their expected biases is given by:
\begin{equation*}
    \mathbb{E}\big[\hat{\tau}^{1}(p, \pi) - \tau\big] - \mathbb{E}\big[\hat{\tau}(p, \pi) - \tau\big] = \frac{1}{n} \sum_{i \neq j} (\beta_{i,j} + p\pi \zeta_{i,j}).
\end{equation*}
This result implies that the absolute bias of $\hat{\tau}(p, \pi)$ is smaller than the absolute bias of $\hat{\tau}^{1}(p, \pi)$. In other words, the estimator $\hat{\tau}(p, \pi)$ is more accurate in estimating the total treatment effect when only a subset of the population is included in the experiment.


\subsection{Sub-population two-stage experiment}
For the remainder of this paper, we will focus on the scenario where the experiment does not include all users in the population. Under a simple Bernoulli randomization experiment without full population coverage, the estimator $\hat{\tau}(p, \pi)$ is biased for the total treatment effect. This naturally leads us to question whether it is possible to find an estimator with a smaller bias. In this subsection, we provide some positive results by considering a special design that is similar to two-stage randomized experiments and the cluster network experiment randomization process \cite{kakhki}. This design aims to reduce bias and improve the accuracy of our estimates, thereby providing a more reliable understanding of the treatment effect.


Given a network graph $G$, let's assume we can partition the graph into disjoint clusters using common clustering algorithm like Louvain community detection \cite{blondel2008fast}. Our two-stage experiment scheme consists of two steps:
\begin{enumerate}
    \item First, we randomly select $p$ percent of the clusters to form the sub-population in the experiment.
    \item Second, we randomly assign the units in the sub-population to either the treatment group with probability $\pi$, or the control group with probability $1-\pi$.
\end{enumerate}

A visualization of this two-stage experiment process is provided in Figure \ref{fig:two_stage_expt}. Similar to the setting in subsection \ref{subsec:subpopulation}, for each user $i$, we use $V_i \in \{0,1\}$ to  indicate whether they are included in the experiment, with $\mathbb{P}(V_i = 1) = p$. Users who are not included in the experiment are assigned the control strategy. To distinguish this scenario from the one discussed in subsection \ref{subsec:subpopulation}, we denote the two HT estimators in \eqref{equa: htate2} and \eqref{equa: htdiffuse2} as $\hat{\tau}^{1}_c(p, \pi)$ and $\hat{\tau}^{2}_c(p, \pi)$, respectively. The HT estimator of the total treatment effect is then given by $\hat{\tau}_c(p, \pi) = \hat{\tau}^{1}_c(p, \pi) + \hat{\tau}^{2}_c(p, \pi)$. Finally, let $\bar{\sigma}$ denote the proportion of each user's friends within the same cluster relative to their total number of friends, i.e.,
\begin{equation*}
    \bar{\sigma} = \frac{1}{n}\sum_{i} \frac{|\{j:j \text{ and }i \text{ are in the same cluster}\}|}{d_i}
\end{equation*}
We have the following results.

\begin{prop}\label{prop: prop6}
$\mathbb{E}\big[\hat{\tau}^{1}_c(p, \pi)\big] = \frac{1}{n} \sum_{i \neq j} \Big(\gamma_{i,j} + \big(\bar{\sigma}\pi+(1-\bar{\sigma})p\pi \big)\zeta_{i,j} \Big) .$
\end{prop}

\begin{prop}\label{prop: prop7}
$\mathbb{E}\big[\hat{\tau}^{2}_c(p, \pi)\big] = \frac{1}{n} \sum_{i \neq j} \Big(\beta_{i,j} + \big(\bar{\sigma}\pi+(1-\bar{\sigma})p\pi \big)\zeta_{i,j} \Big) .$
\end{prop}

\begin{theorem}\label{thm: thm3}
Assume (A\ref{hyp:pair}). Under a two-stage experiment scheme, the expectation of $\hat{\tau}_c(p, \pi)$ is
\begin{equation}
    \mathbb{E}\big[\hat{\tau}_c(p, \pi)\big] =  \frac{1}{n} \sum_{i \neq j} \Big(\beta_{i,j} + \gamma_{i,j} + 2\pi\big(\bar{\sigma}+(1-\bar{\sigma})p \big)\zeta_{i,j} \Big).
\end{equation}
The bias is 
\begin{equation}\label{equa:bias_cluster_sub}
    \mathbb{E}\big[\hat{\tau}^{2}_c(p, \pi) -\tau \big] = - \frac{(1-2p\pi)(1-\bar{\sigma})}{n} \sum_{i \neq j}\zeta_{i,j}.
\end{equation}
\end{theorem}

From equations \eqref{equa:bias_sub} and \eqref{equa:bias_cluster_sub}, we can observe that the absolute bias of the estimator $\hat{\tau}_c(p, \pi)$ in the two-stage experiment is reduced by $\frac{\bar{\sigma}(1-2p\pi)}{n} \sum_{i \neq j}\zeta_{i,j}$ compared to the absolute bias of the estimator $\hat{\tau}(p, \pi)$ in the simple randomized experiment. This indicates that the two-stage experiment can further reduce the bias of the estimator without sacrificing any statistical power. This is a significant finding, as it suggests that a two-stage experimental design can provide more accurate estimates of the treatment effect, thereby improving the reliability of our results. In practical applications, the value of $\bar{\sigma}$ typically falls between  $30\%$ to $40\%$ \cite{kakhki}, meaning that the bias can be reduced by a minimum of 30\%.

\section{Numerical Experiments}\label{sec4}

\subsection{Simulated data}
To compare the performance of the estimators mentioned earlier under various conditions, we conducted several simulation studies. These studies were based on two real-world social networks: FB-Standard3 and FB-Cornell5 \footnote{Network topology data can be found in https://networkrepository.com/socfb-Stanford3.php and https://networkrepository.com/socfb-Cornell5.php}. The FB-Standard3 network comprises 11,586 nodes and 568,309 edges, yielding an average degree of 98.1. On the other hand, the FB-Cornell5 network consists of 18,660 nodes and 790,777 edges, resulting in an average degree of 84.7. To initiate our simulations, we first generated the paired outcomes based on the model defined in equation \eqref{equa: zijmodel}:
\begin{equation*}
    z_{i,j}(W_i,W_j) = \alpha_{i,j} + \beta_{i, j} W_i + \gamma_{i, j} W_j + \zeta_{i,j} W_i W_j.
\end{equation*}
Next, we aggregated the paired outcomes to obtain $Y_i$ and $D_i$. The parameters $\alpha_{i,j}, \beta_{i,j}, \gamma_{i,j}, \zeta_{i,j}$ were generated from either a uniform or a Bernoulli distribution. Specifically, we considered the following two cases:
\begin{itemize}
    \item For uniform distributed coefficients, $\alpha_{i,j} =1$, $\beta_{i,j} \sim $ Uniform $(0, 0.5)$, $\gamma_{i,j} \sim \text{Uniform}(0, 1)$, $\zeta_{i,j} \sim \text{Uniform} $ $(0, 0.5)$ for all pair $(i,j) \in E$ independently.
    \item For Bernoulli distributed coefficients, $\alpha_{i,j} = 0$, $\beta_{i,j} \sim $ Bernoulli $(0.25)$, $\gamma_{i,j} \sim \text{Bernoulli}(0.5)$, $\zeta_{i,j} \sim \text{Bernoulli}(0.25)$ for all pair $(i,j) \in E$ independently.
\end{itemize}
Once the parameters were generated, they remained fixed throughout the simulation. We considered a total of four different simulation settings. For each setting, we performed 10,000 repetitions to calculate the bias of the estimators and assess their performance.

\begin{figure*}[htbp]
\centering
\includegraphics[width=1\textwidth]{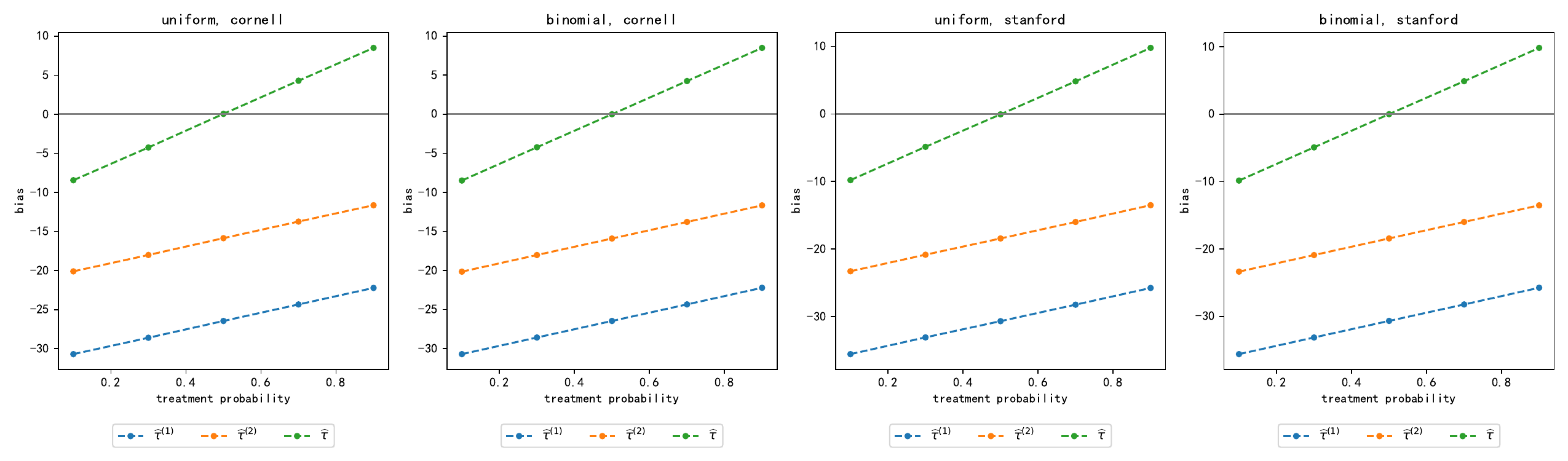}
\caption{
Visualizations of the performance of three TTE estimators under full population Bernoulli design on FB-Standard3 and FB-Cornell5 networks for both Uniform and Bernoulli potential outcomes models. Each line's height represents the relative bias of the estimator.
}
\label{fig:simulation_all_population}
\end{figure*}

Firstly, we simulated a scenario where the entire population is included in the experiment. We varied the treatment probability $\pi$ across the set $\{0.1, 0.3, 0.5, 0.7,$ $0.9\}$. The results of these simulations are presented in Figure \ref{fig:simulation_all_population}. From the figure, it is evident that both $\hat{\tau}^1(\pi)$ and $\hat{\tau}^2(\pi)$ exhibit bias under any given treatment probability. However, $\hat{\tau}(\pi)$ remains unbiased when $\pi=0.5$. This observation aligns perfectly with the conclusion drawn in Theorem \ref{thm: thm1}.

\begin{figure*}[htbp]
\centering
\includegraphics[width=1\textwidth]{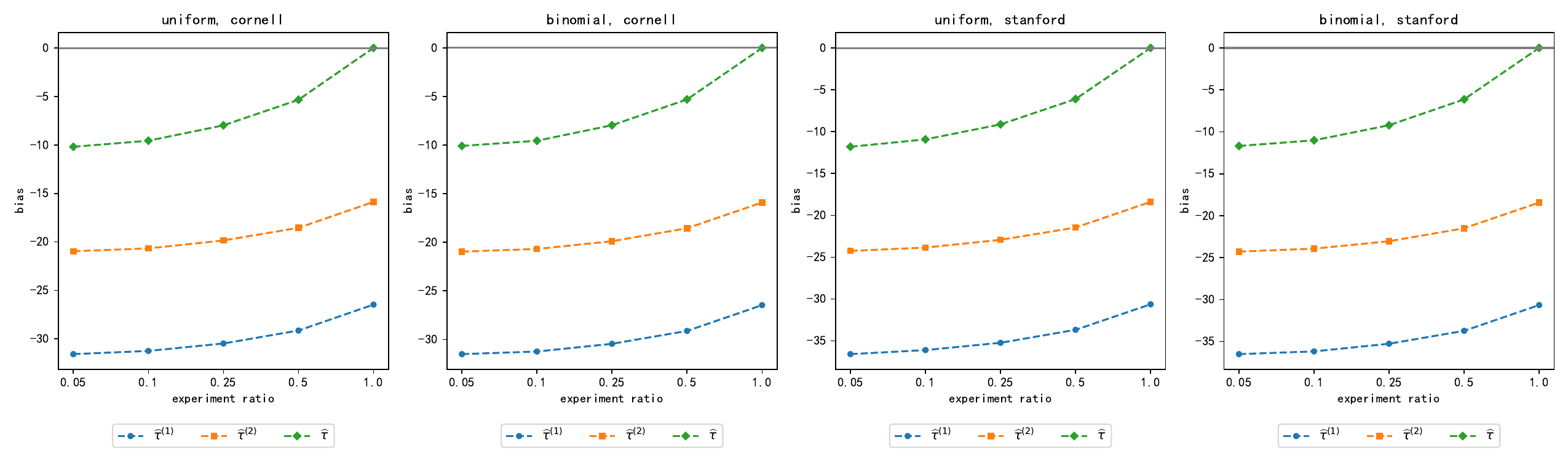}
\caption{
Visualizations of the performance of three TTE estimators under sub-population Bernoulli design on FB-Standard3 and FB-Cornell5 networks for both Uniform and Bernoulli potential outcomes models. Each line's height represents the relative bias of the estimator.
}
\label{fig:simulation_sub_population}
\end{figure*}

Secondly, we simulated a scenario where only a subset of the population is included in the experiment. In this case, we fixed the treatment probability at $\pi=0.5$, varied the in-experiment probability $p$ across the set $\{0.05, 0.1, 0.25, 0.5, 1\}$. The results of these simulations are presented in Figure \ref{fig:simulation_sub_population}. From the figure, it is evident that when $p < 1$ , all three estimators $\hat{\tau}^1(p, \pi)$, $\hat{\tau}^2(p, \pi)$ and $\hat{\tau}(p, \pi)$ exhibit bias. Not surprisingly, the bias of $\hat{\tau}(p, \pi)$ decreases as $p$ approaches 1. This suggests that increasing the traffic ratio of the experiment can help reduce the bias introduced by interference. Furthermore, we observe that the absolute bias of $\hat{\tau}(p, \pi)$ is consistently smaller than the absolute bias of $\hat{\tau}^1(p, \pi)$ under any given in-experiment probability. This indicates the superiority of $\hat{\tau}(p, \pi)$ over $\hat{\tau}^1(p, \pi)$.

\begin{figure*}[htbp]
\centering
\includegraphics[width=1\textwidth]{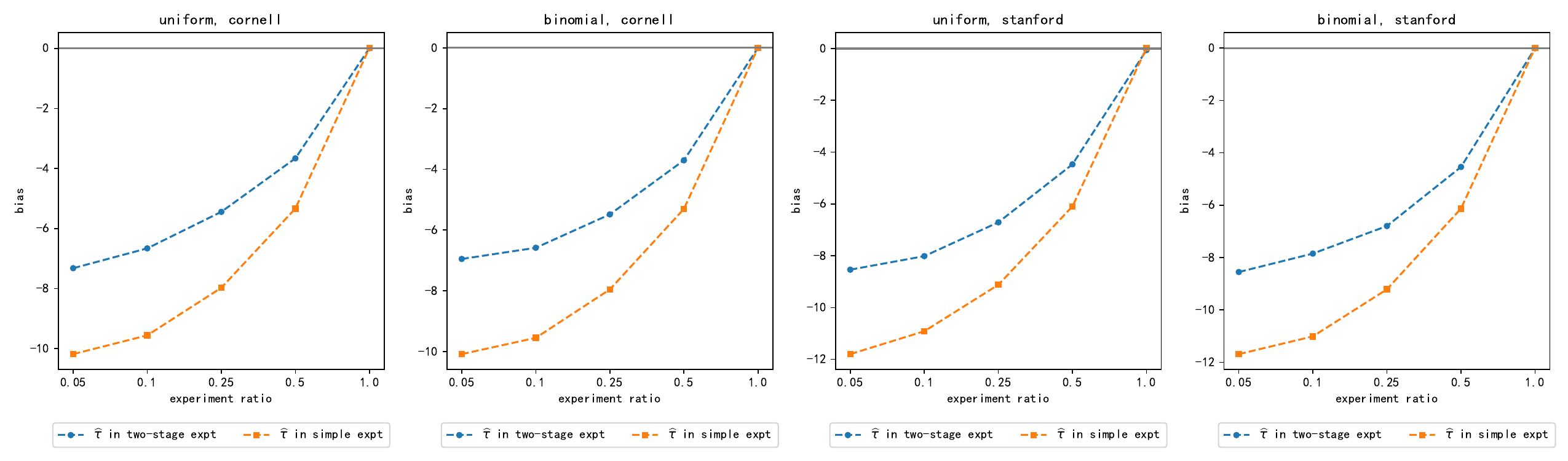}
\caption{
Visualizations of the performance of three TTE estimators under sub-population two-stage experiment design on FB-Standard3 and FB-Cornell5 networks for both Uniform and Bernoulli potential outcomes models. Each line's height represents the relative bias of the estimator.
}
\label{fig:simulation_cluster_sub_population}
\end{figure*}

Lastly, we simulated a two-stage experiment scenario. We first applied the Louvain algorithm \cite{blondel2008fast} with a fixed random seed and a resolution parameter set to 10 to group the units into clusters. This resulted in 240 clusters for the FB-Cornell5 graph and 192 clusters for the FB-Stanford3 graph. As in the previous scenario, we fixed the treatment probability at $\pi=0.5$, and varied the in-experiment probability $p$ across the set $\{0.05, 0.1, 0.25, 0.5, 1\}$. The results of these simulations are presented in Figure \ref{fig:simulation_cluster_sub_population}. From the figure, it is evident that the two-stage experiment scheme can further reduce the bias of $\hat{\tau}(p, \pi)$ compared to the simple Bernoulli randomized experiment. This suggests that a two-stage experimental design, where units are first grouped into clusters, can be a more effective strategy for reducing bias in the presence of interference.

\subsection{Real Data}

\begin{table*}[htbp]
\caption{Estimates with a re-scaled confidence interval of the metrics in the two-stage experiment.}\label{tab1}
\begin{center}
\setlength{\tabcolsep}{5mm}{
\begin{tabular}{l|c|c|c}
  \toprule
   & $\hat{\tau}^1$ & $\hat{\tau}^2$ & $\hat{\tau}$  \\
  \hline
  Metric 1 & -0.605\% $\pm$ 0.255\% & 3.498\% $\pm$ 1.198\% & 0.621\% $\pm$ 0.5\%   \\
  Metric 2 & -0.712\% $\pm$ 0.349\% & 3.381\% $\pm$ 1.171\% & 0.678\% $\pm$ 0.5\% \\
  \bottomrule
\end{tabular}
}
\end{center}
\end{table*}

We present an empirical experiment conducted on Weixin's experiment platform, utilizing aggregated dyadic data to estimate the total treatment effect. Weixin, a popular messaging and calling application, allows users to interact with their contacts. A built-in feature of Weixin, known as "Channels", is akin to TikTok, enabling users to view and discover a plethora of personalized short videos. Similar to Example \ref{exmp:pair}, users typically share intriguing videos with their friends, who can then access the "Channels" page by clicking on the shared video, watching it, and exploring other videos. 

A two-stage experiment spanning two weeks was executed to evaluate a novel algorithm. The experiment contains about 4\% of the users. In the treatment group, the algorithm increased the likelihood of users sharing videos. The underlying hypothesis was that promoting video sharing would lead to users exiting the "Channels" page to engage in conversations with friends, thereby reducing their consumption but increasing that of their friends. The primary objective was to ascertain whether this new algorithm could enhance user activity and consumption when deployed to all users.

The outcomes for two crucial metrics are detailed in Table \ref{tab1}. All data and attributes are collected after user approvals and data masking to protect user privacy. Each metric was re-scaled such that the estimator $\hat{\tau}$ has a confidence interval (CI) width of one. The results align with our initial expectations. The $\hat{\tau}^1$ indicates a significant decrease, while the $\hat{\tau}$ exhibits a substantial increase, implying that the total treatment effect of the new algorithm is positive. Upon deploying the new algorithm to all users, an upward trend in the consumption of the entire user base was observed, corroborating the results of $\hat{\tau}$.

\section{Discussion}\label{sec5}
The estimator $\hat{\tau}$, as developed in our study, has the potential to estimate the total treatment effect within a dyadic data structure. This estimator exhibits no bias when the experiment encompasses the entire population and the probability of treatment is set at 0.5. Compared to the conventional estimator $\hat{\tau}^1$, our proposed estimator $\hat{\tau}$ demonstrates reduced bias when the experiment includes only a small subset of the population. Furthermore, we suggest the application of $\hat{\tau}$ within a two-stage experimental scheme, a strategy that can further mitigate the estimator's bias.

Several potential avenues exist for extending our current work. Firstly, we have not yet presented the variance properties of these estimators. In practical applications, we utilize the sample variance of $Y_i$ and $D_i$ to facilitate statistical inference. Our simulation findings suggest that the sample variance provides a reasonably accurate estimate of the true variance. Future work could involve deriving the expression for the variance of these estimators, demonstrating the convergence rate, and proposing alternative estimators for variance within a dyadic data structure.

Secondly, our work relies on an important assumption (A\ref{hyp:pair}), which posits that the potential pair outcomes $z_{i,j}(\mathbf{W})$ depend solely on the treatment assignments of units $i$ and $j$. This assumption could be extended to a more general case, for instance, by assuming that the potential pair outcomes $z_{i,j}(\mathbf{W})$ depend on the treatment assignments of units $i, j$, and all their respective neighbors. Although an assumption of this nature significantly increases the complexity of inferring the total treatment effect, researchers may be able to leverage dyadic data to construct estimators that exhibit superior properties compared to commonly used estimators, such as the difference-in-means estimator and the Horvitz-Thompson estimator.

Thirdly, this paper considers simple Bernoulli randomization experimentation. There are other experimental randomization schemes, such as cluster randomization or ego cluster randomization, which are useful for dealing with interference. Future research could explore the integration of our estimators with these alternative randomization schemes, potentially leading to insightful findings for estimating the total treatment effect.

Fourthly, dyadic data represents a unique type of data structure. However, in some scenarios, we may not have access to dyadic data, meaning that the outcome $Y_i$ cannot be decomposed into the sum of edge-level outcomes. Consider, for instance, a user activity level metric, which is binary, represented by 1 (active) or 0 (inactive). It is not feasible to express the user activity level metric in a summation form. For such metrics, researchers could consider integrating our method with the approaches proposed in \cite{li2022random, hu2022average} to estimate the total treatment effect.

Exploring these potential extensions could significantly broaden the applicability of our estimators to more general models, while also providing the necessary theoretical justifications. These promising research directions present exciting opportunities for future work, and we look forward to seeing how they will further advance our understanding in this field.
\\

\noindent\textbf{Code Availability Statement:} 
\href{https://github.com/adamdenglu/diffusion-metric-with-aggregated-dyadic-data}{https://github.com/adamdenglu/diffusion-metric-with-aggregated-dyadic-data}


\bibliographystyle{ACM-Reference-Format}
\bibliography{ref}


\begin{thebibliography}{39}


\ifx \showCODEN    \undefined \def \showCODEN     #1{\unskip}     \fi
\ifx \showDOI      \undefined \def \showDOI       #1{#1}\fi
\ifx \showISBNx    \undefined \def \showISBNx     #1{\unskip}     \fi
\ifx \showISBNxiii \undefined \def \showISBNxiii  #1{\unskip}     \fi
\ifx \showISSN     \undefined \def \showISSN      #1{\unskip}     \fi
\ifx \showLCCN     \undefined \def \showLCCN      #1{\unskip}     \fi
\ifx \shownote     \undefined \def \shownote      #1{#1}          \fi
\ifx \showarticletitle \undefined \def \showarticletitle #1{#1}   \fi
\ifx \showURL      \undefined \def \showURL       {\relax}        \fi
\providecommand\bibfield[2]{#2}
\providecommand\bibinfo[2]{#2}
\providecommand\natexlab[1]{#1}
\providecommand\showeprint[2][]{arXiv:#2}

\bibitem[Apicella et~al\mbox{.}(2012)]%
        {apicella2012}
\bibfield{author}{\bibinfo{person}{Coren~L Apicella}, \bibinfo{person}{Frank~W
  Marlowe}, \bibinfo{person}{James~H Fowler}, {and} \bibinfo{person}{Nicholas~A
  Christakis}.} \bibinfo{year}{2012}\natexlab{}.
\newblock \showarticletitle{Social networks and cooperation in
  hunter-gatherers}.
\newblock \bibinfo{journal}{\emph{Nature}} \bibinfo{volume}{481},
  \bibinfo{number}{7382} (\bibinfo{year}{2012}), \bibinfo{pages}{497--501}.
\newblock


\bibitem[Aronow and Samii(2012)]%
        {aronow2012}
\bibfield{author}{\bibinfo{person}{Peter~M Aronow} {and} \bibinfo{person}{Cyrus
  Samii}.} \bibinfo{year}{2012}\natexlab{}.
\newblock \showarticletitle{Estimating average causal effects under general
  interference}. In \bibinfo{booktitle}{\emph{Summer Meeting of the Society for
  Political Methodology, University of North Carolina, Chapel Hill, July}}.
  \bibinfo{pages}{19--21}.
\newblock


\bibitem[Aronow and Samii(2017)]%
        {aronow2017estimating}
\bibfield{author}{\bibinfo{person}{Peter~M Aronow} {and} \bibinfo{person}{Cyrus
  Samii}.} \bibinfo{year}{2017}\natexlab{}.
\newblock \showarticletitle{Estimating average causal effects under general
  interference, with application to a social network experiment}.
\newblock  (\bibinfo{year}{2017}).
\newblock


\bibitem[Athey et~al\mbox{.}(2018)]%
        {athey}
\bibfield{author}{\bibinfo{person}{Susan Athey}, \bibinfo{person}{Dean Eckles},
  {and} \bibinfo{person}{Guido~W Imbens}.} \bibinfo{year}{2018}\natexlab{}.
\newblock \showarticletitle{Exact p-values for network interference}.
\newblock \bibinfo{journal}{\emph{J. Amer. Statist. Assoc.}}
  \bibinfo{volume}{113}, \bibinfo{number}{521} (\bibinfo{year}{2018}),
  \bibinfo{pages}{230--240}.
\newblock


\bibitem[Basse and Airoldi(2015)]%
        {basse2015optimal}
\bibfield{author}{\bibinfo{person}{Guillaume~W Basse} {and}
  \bibinfo{person}{Edoardo~M Airoldi}.} \bibinfo{year}{2015}\natexlab{}.
\newblock \showarticletitle{Optimal model-assisted design of experiments for
  network correlated outcomes suggests new notions of network balance}.
\newblock \bibinfo{journal}{\emph{arXiv preprint arXiv:1507.00803}}
  (\bibinfo{year}{2015}).
\newblock


\bibitem[Basse and Airoldi(2018)]%
        {basse2017}
\bibfield{author}{\bibinfo{person}{Guillaume~W Basse} {and}
  \bibinfo{person}{Edoardo~M Airoldi}.} \bibinfo{year}{2018}\natexlab{}.
\newblock \showarticletitle{Model-assisted design of experiments in the
  presence of network-correlated outcomes}.
\newblock \bibinfo{journal}{\emph{Biometrika}} \bibinfo{volume}{105},
  \bibinfo{number}{4} (\bibinfo{year}{2018}), \bibinfo{pages}{849--858}.
\newblock


\bibitem[Blondel et~al\mbox{.}(2008)]%
        {blondel2008fast}
\bibfield{author}{\bibinfo{person}{Vincent~D Blondel},
  \bibinfo{person}{Jean-Loup Guillaume}, \bibinfo{person}{Renaud Lambiotte},
  {and} \bibinfo{person}{Etienne Lefebvre}.} \bibinfo{year}{2008}\natexlab{}.
\newblock \showarticletitle{Fast unfolding of communities in large networks}.
\newblock \bibinfo{journal}{\emph{Journal of statistical mechanics: theory and
  experiment}} \bibinfo{volume}{2008}, \bibinfo{number}{10}
  (\bibinfo{year}{2008}), \bibinfo{pages}{P10008}.
\newblock


\bibitem[Chin(2018)]%
        {chin2018}
\bibfield{author}{\bibinfo{person}{Alex Chin}.}
  \bibinfo{year}{2018}\natexlab{}.
\newblock \showarticletitle{Central limit theorems via Stein's method for
  randomized experiments under interference}.
\newblock \bibinfo{journal}{\emph{arXiv preprint arXiv:1804.03105}}
  (\bibinfo{year}{2018}).
\newblock


\bibitem[Cortez et~al\mbox{.}(2022)]%
        {cortez2022staggered}
\bibfield{author}{\bibinfo{person}{Mayleen Cortez}, \bibinfo{person}{Matthew
  Eichhorn}, {and} \bibinfo{person}{Christina Yu}.}
  \bibinfo{year}{2022}\natexlab{}.
\newblock \showarticletitle{Staggered rollout designs enable causal inference
  under interference without network knowledge}.
\newblock \bibinfo{journal}{\emph{Advances in Neural Information Processing
  Systems}}  \bibinfo{volume}{35} (\bibinfo{year}{2022}),
  \bibinfo{pages}{7437--7449}.
\newblock


\bibitem[Eckles et~al\mbox{.}(2016)]%
        {eckles}
\bibfield{author}{\bibinfo{person}{Dean Eckles}, \bibinfo{person}{Brian
  Karrer}, {and} \bibinfo{person}{Johan Ugander}.}
  \bibinfo{year}{2016}\natexlab{}.
\newblock \showarticletitle{Design and analysis of experiments in networks:
  Reducing bias from interference}.
\newblock \bibinfo{journal}{\emph{Journal of Causal Inference}}
  \bibinfo{volume}{5}, \bibinfo{number}{1} (\bibinfo{year}{2016}),
  \bibinfo{pages}{20150021}.
\newblock


\bibitem[Forastiere et~al\mbox{.}(2021)]%
        {forastiere2021identification}
\bibfield{author}{\bibinfo{person}{Laura Forastiere},
  \bibinfo{person}{Edoardo~M Airoldi}, {and} \bibinfo{person}{Fabrizia
  Mealli}.} \bibinfo{year}{2021}\natexlab{}.
\newblock \showarticletitle{Identification and estimation of treatment and
  interference effects in observational studies on networks}.
\newblock \bibinfo{journal}{\emph{J. Amer. Statist. Assoc.}}
  \bibinfo{volume}{116}, \bibinfo{number}{534} (\bibinfo{year}{2021}),
  \bibinfo{pages}{901--918}.
\newblock


\bibitem[Forastiere et~al\mbox{.}(2022)]%
        {forastiere2022estimating}
\bibfield{author}{\bibinfo{person}{Laura Forastiere}, \bibinfo{person}{Fabrizia
  Mealli}, \bibinfo{person}{Albert Wu}, {and} \bibinfo{person}{Edoardo~M
  Airoldi}.} \bibinfo{year}{2022}\natexlab{}.
\newblock \showarticletitle{Estimating causal effects under network
  interference with Bayesian generalized propensity scores}.
\newblock \bibinfo{journal}{\emph{The Journal of Machine Learning Research}}
  \bibinfo{volume}{23}, \bibinfo{number}{1} (\bibinfo{year}{2022}),
  \bibinfo{pages}{13101--13161}.
\newblock


\bibitem[Gui et~al\mbox{.}(2015)]%
        {gui}
\bibfield{author}{\bibinfo{person}{Huan Gui}, \bibinfo{person}{Ya Xu},
  \bibinfo{person}{Anmol Bhasin}, {and} \bibinfo{person}{Jiawei Han}.}
  \bibinfo{year}{2015}\natexlab{}.
\newblock \showarticletitle{Network a/b testing: From sampling to estimation}.
  In \bibinfo{booktitle}{\emph{Proceedings of the 24th International Conference
  on World Wide Web}}. \bibinfo{pages}{399--409}.
\newblock


\bibitem[Hayes and Moulton(2017)]%
        {hayes}
\bibfield{author}{\bibinfo{person}{Richard~J Hayes} {and}
  \bibinfo{person}{Lawrence~H Moulton}.} \bibinfo{year}{2017}\natexlab{}.
\newblock \bibinfo{booktitle}{\emph{Cluster randomised trials}}.
\newblock \bibinfo{publisher}{CRC press}.
\newblock


\bibitem[Hu et~al\mbox{.}(2022)]%
        {hu2022average}
\bibfield{author}{\bibinfo{person}{Yuchen Hu}, \bibinfo{person}{Shuangning Li},
  {and} \bibinfo{person}{Stefan Wager}.} \bibinfo{year}{2022}\natexlab{}.
\newblock \showarticletitle{Average direct and indirect causal effects under
  interference}.
\newblock \bibinfo{journal}{\emph{Biometrika}} \bibinfo{volume}{109},
  \bibinfo{number}{4} (\bibinfo{year}{2022}), \bibinfo{pages}{1165--1172}.
\newblock


\bibitem[Imbens and Rubin(2015)]%
        {imbens}
\bibfield{author}{\bibinfo{person}{Guido~W Imbens} {and}
  \bibinfo{person}{Donald~B Rubin}.} \bibinfo{year}{2015}\natexlab{}.
\newblock \bibinfo{booktitle}{\emph{Causal inference in statistics, social, and
  biomedical sciences}}.
\newblock \bibinfo{publisher}{Cambridge University Press}.
\newblock


\bibitem[Karrer et~al\mbox{.}(2021)]%
        {kakhki}
\bibfield{author}{\bibinfo{person}{Brian Karrer}, \bibinfo{person}{Liang Shi},
  \bibinfo{person}{Monica Bhole}, \bibinfo{person}{Matt Goldman},
  \bibinfo{person}{Tyrone Palmer}, \bibinfo{person}{Charlie Gelman},
  \bibinfo{person}{Mikael Konutgan}, {and} \bibinfo{person}{Feng Sun}.}
  \bibinfo{year}{2021}\natexlab{}.
\newblock \showarticletitle{Network experimentation at scale}. In
  \bibinfo{booktitle}{\emph{Proceedings of the 27th acm sigkdd conference on
  knowledge discovery \& data mining}}. \bibinfo{pages}{3106--3116}.
\newblock


\bibitem[Kohavi et~al\mbox{.}(2013)]%
        {kohavi1}
\bibfield{author}{\bibinfo{person}{Ron Kohavi}, \bibinfo{person}{Alex Deng},
  \bibinfo{person}{Brian Frasca}, \bibinfo{person}{Toby Walker},
  \bibinfo{person}{Ya Xu}, {and} \bibinfo{person}{Nils Pohlmann}.}
  \bibinfo{year}{2013}\natexlab{}.
\newblock \showarticletitle{Online controlled experiments at large scale}. In
  \bibinfo{booktitle}{\emph{Proceedings of the 19th ACM SIGKDD international
  conference on Knowledge discovery and data mining}}.
  \bibinfo{pages}{1168--1176}.
\newblock


\bibitem[Kohavi et~al\mbox{.}(2014)]%
        {kohavi2}
\bibfield{author}{\bibinfo{person}{Ron Kohavi}, \bibinfo{person}{Alex Deng},
  \bibinfo{person}{Roger Longbotham}, {and} \bibinfo{person}{Ya Xu}.}
  \bibinfo{year}{2014}\natexlab{}.
\newblock \showarticletitle{Seven rules of thumb for web site experimenters}.
  In \bibinfo{booktitle}{\emph{Proceedings of the 20th ACM SIGKDD international
  conference on Knowledge discovery and data mining}}.
  \bibinfo{pages}{1857--1866}.
\newblock


\bibitem[Leung(2022)]%
        {leung2022causal}
\bibfield{author}{\bibinfo{person}{Michael~P Leung}.}
  \bibinfo{year}{2022}\natexlab{}.
\newblock \showarticletitle{Causal inference under approximate neighborhood
  interference}.
\newblock \bibinfo{journal}{\emph{Econometrica}} \bibinfo{volume}{90},
  \bibinfo{number}{1} (\bibinfo{year}{2022}), \bibinfo{pages}{267--293}.
\newblock


\bibitem[Li and Wager(2022)]%
        {li2022random}
\bibfield{author}{\bibinfo{person}{Shuangning Li} {and} \bibinfo{person}{Stefan
  Wager}.} \bibinfo{year}{2022}\natexlab{}.
\newblock \showarticletitle{Random graph asymptotics for treatment effect
  estimation under network interference}.
\newblock \bibinfo{journal}{\emph{The Annals of Statistics}}
  \bibinfo{volume}{50}, \bibinfo{number}{4} (\bibinfo{year}{2022}),
  \bibinfo{pages}{2334--2358}.
\newblock


\bibitem[Liu et~al\mbox{.}(2022)]%
        {liu2022adaptive}
\bibfield{author}{\bibinfo{person}{Yang Liu}, \bibinfo{person}{Yifan Zhou},
  \bibinfo{person}{Ping Li}, {and} \bibinfo{person}{Feifang Hu}.}
  \bibinfo{year}{2022}\natexlab{}.
\newblock \showarticletitle{Adaptive A/B Test on Networks with Cluster
  Structures}. In \bibinfo{booktitle}{\emph{International Conference on
  Artificial Intelligence and Statistics}}. PMLR,
  \bibinfo{pages}{10836--10851}.
\newblock


\bibitem[Luke and Harris(2007)]%
        {luke2007}
\bibfield{author}{\bibinfo{person}{Douglas~A Luke} {and}
  \bibinfo{person}{Jenine~K Harris}.} \bibinfo{year}{2007}\natexlab{}.
\newblock \showarticletitle{Network analysis in public health: history,
  methods, and applications}.
\newblock \bibinfo{journal}{\emph{Annu. Rev. Public Health}}
  \bibinfo{volume}{28} (\bibinfo{year}{2007}), \bibinfo{pages}{69--93}.
\newblock


\bibitem[Manski(2013)]%
        {manski2013}
\bibfield{author}{\bibinfo{person}{Charles~F Manski}.}
  \bibinfo{year}{2013}\natexlab{}.
\newblock \showarticletitle{Identification of treatment response with social
  interactions}.
\newblock \bibinfo{journal}{\emph{The Econometrics Journal}}
  \bibinfo{volume}{16}, \bibinfo{number}{1} (\bibinfo{year}{2013}),
  \bibinfo{pages}{S1--S23}.
\newblock


\bibitem[Portes and Rey(2005)]%
        {portes2005}
\bibfield{author}{\bibinfo{person}{Richard Portes} {and}
  \bibinfo{person}{Helene Rey}.} \bibinfo{year}{2005}\natexlab{}.
\newblock \showarticletitle{The determinants of cross-border equity flows}.
\newblock \bibinfo{journal}{\emph{Journal of international Economics}}
  \bibinfo{volume}{65}, \bibinfo{number}{2} (\bibinfo{year}{2005}),
  \bibinfo{pages}{269--296}.
\newblock


\bibitem[Pouget-Abadie et~al\mbox{.}(2019)]%
        {abadie2019}
\bibfield{author}{\bibinfo{person}{Jean Pouget-Abadie},
  \bibinfo{person}{Guillaume Saint-Jacques}, \bibinfo{person}{Martin Saveski},
  \bibinfo{person}{Weitao Duan}, \bibinfo{person}{Souvik Ghosh},
  \bibinfo{person}{Ya Xu}, {and} \bibinfo{person}{Edoardo~M Airoldi}.}
  \bibinfo{year}{2019}\natexlab{}.
\newblock \showarticletitle{Testing for arbitrary interference on
  experimentation platforms}.
\newblock \bibinfo{journal}{\emph{Biometrika}} \bibinfo{volume}{106},
  \bibinfo{number}{4} (\bibinfo{year}{2019}), \bibinfo{pages}{929--940}.
\newblock


\bibitem[Saint-Jacques et~al\mbox{.}(2019a)]%
        {saint2019method}
\bibfield{author}{\bibinfo{person}{Guillaume Saint-Jacques},
  \bibinfo{person}{James~Eric Sorenson}, \bibinfo{person}{Nanyu Chen}, {and}
  \bibinfo{person}{Ya Xu}.} \bibinfo{year}{2019}\natexlab{a}.
\newblock \showarticletitle{A method for measuring network effects of
  one-to-one communication features in online a/b tests}.
\newblock \bibinfo{journal}{\emph{arXiv preprint arXiv:1903.08766}}
  (\bibinfo{year}{2019}).
\newblock


\bibitem[Saint-Jacques et~al\mbox{.}(2019b)]%
        {saintjacques}
\bibfield{author}{\bibinfo{person}{Guillaume Saint-Jacques},
  \bibinfo{person}{Maneesh Varshney}, \bibinfo{person}{Jeremy Simpson}, {and}
  \bibinfo{person}{Ya Xu}.} \bibinfo{year}{2019}\natexlab{b}.
\newblock \showarticletitle{Using ego-clusters to measure network effects at
  LinkedIn}.
\newblock \bibinfo{journal}{\emph{arXiv preprint arXiv:1903.08755}}
  (\bibinfo{year}{2019}).
\newblock


\bibitem[S{\"a}rndal et~al\mbox{.}(2003)]%
        {sarndal2003}
\bibfield{author}{\bibinfo{person}{Carl-Erik S{\"a}rndal},
  \bibinfo{person}{Bengt Swensson}, {and} \bibinfo{person}{Jan Wretman}.}
  \bibinfo{year}{2003}\natexlab{}.
\newblock \bibinfo{booktitle}{\emph{Model assisted survey sampling}}.
\newblock \bibinfo{publisher}{Springer Science \& Business Media}.
\newblock


\bibitem[S{\"a}vje(2023)]%
        {savje2023causal}
\bibfield{author}{\bibinfo{person}{Fredrik S{\"a}vje}.}
  \bibinfo{year}{2023}\natexlab{}.
\newblock \showarticletitle{Causal inference with misspecified exposure
  mappings: separating definitions and assumptions}.
\newblock \bibinfo{journal}{\emph{Biometrika}} (\bibinfo{year}{2023}),
  \bibinfo{pages}{asad019}.
\newblock


\bibitem[S{\"a}vje et~al\mbox{.}(2021)]%
        {savje2021average}
\bibfield{author}{\bibinfo{person}{Fredrik S{\"a}vje}, \bibinfo{person}{Peter
  Aronow}, {and} \bibinfo{person}{Michael Hudgens}.}
  \bibinfo{year}{2021}\natexlab{}.
\newblock \showarticletitle{Average treatment effects in the presence of
  unknown interference}.
\newblock \bibinfo{journal}{\emph{Annals of statistics}} \bibinfo{volume}{49},
  \bibinfo{number}{2} (\bibinfo{year}{2021}), \bibinfo{pages}{673}.
\newblock


\bibitem[Sussman and Airoldi(2017)]%
        {sussman2017elements}
\bibfield{author}{\bibinfo{person}{Daniel~L Sussman} {and}
  \bibinfo{person}{Edoardo~M Airoldi}.} \bibinfo{year}{2017}\natexlab{}.
\newblock \showarticletitle{Elements of estimation theory for causal effects in
  the presence of network interference}.
\newblock \bibinfo{journal}{\emph{arXiv preprint arXiv:1702.03578}}
  (\bibinfo{year}{2017}).
\newblock


\bibitem[Tang et~al\mbox{.}(2010)]%
        {tang2010}
\bibfield{author}{\bibinfo{person}{Diane Tang}, \bibinfo{person}{Ashish
  Agarwal}, \bibinfo{person}{Deirdre O'Brien}, {and} \bibinfo{person}{Mike
  Meyer}.} \bibinfo{year}{2010}\natexlab{}.
\newblock \showarticletitle{Overlapping experiment infrastructure: More,
  better, faster experimentation}. In \bibinfo{booktitle}{\emph{Proceedings of
  the 16th ACM SIGKDD international conference on Knowledge discovery and data
  mining}}. \bibinfo{pages}{17--26}.
\newblock


\bibitem[Toulis and Kao(2013)]%
        {toulis2013}
\bibfield{author}{\bibinfo{person}{Panos Toulis} {and} \bibinfo{person}{Edward
  Kao}.} \bibinfo{year}{2013}\natexlab{}.
\newblock \showarticletitle{Estimation of causal peer influence effects}. In
  \bibinfo{booktitle}{\emph{International conference on machine learning}}.
  PMLR, \bibinfo{pages}{1489--1497}.
\newblock


\bibitem[Ugander et~al\mbox{.}(2013)]%
        {ugander}
\bibfield{author}{\bibinfo{person}{Johan Ugander}, \bibinfo{person}{Brian
  Karrer}, \bibinfo{person}{Lars Backstrom}, {and} \bibinfo{person}{Jon
  Kleinberg}.} \bibinfo{year}{2013}\natexlab{}.
\newblock \showarticletitle{Graph cluster randomization: Network exposure to
  multiple universes}. In \bibinfo{booktitle}{\emph{Proceedings of the 19th ACM
  SIGKDD international conference on Knowledge discovery and data mining}}.
  \bibinfo{pages}{329--337}.
\newblock


\bibitem[Ugander and Yin(2023)]%
        {Ugander2023}
\bibfield{author}{\bibinfo{person}{Johan Ugander} {and} \bibinfo{person}{Hao
  Yin}.} \bibinfo{year}{2023}\natexlab{}.
\newblock \showarticletitle{Randomized graph cluster randomization}.
\newblock \bibinfo{journal}{\emph{Journal of Causal Inference}}
  \bibinfo{volume}{11}, \bibinfo{number}{1} (\bibinfo{year}{2023}),
  \bibinfo{pages}{20220014}.
\newblock


\bibitem[Xiong et~al\mbox{.}(2020)]%
        {xiong2020}
\bibfield{author}{\bibinfo{person}{Tao Xiong}, \bibinfo{person}{Yong Wang},
  {and} \bibinfo{person}{Senlie Zheng}.} \bibinfo{year}{2020}\natexlab{}.
\newblock \bibinfo{booktitle}{\emph{Orthogonal Traffic Assignment in Online
  Overlapping A/B Tests}}.
\newblock \bibinfo{type}{{T}echnical {R}eport}.
  \bibinfo{institution}{EasyChair}.
\newblock


\bibitem[Yu et~al\mbox{.}(2022)]%
        {yu2022estimating}
\bibfield{author}{\bibinfo{person}{Christina~Lee Yu},
  \bibinfo{person}{Edoardo~M Airoldi}, \bibinfo{person}{Christian Borgs}, {and}
  \bibinfo{person}{Jennifer~T Chayes}.} \bibinfo{year}{2022}\natexlab{}.
\newblock \showarticletitle{Estimating the total treatment effect in randomized
  experiments with unknown network structure}.
\newblock \bibinfo{journal}{\emph{Proceedings of the National Academy of
  Sciences}} \bibinfo{volume}{119}, \bibinfo{number}{44}
  (\bibinfo{year}{2022}), \bibinfo{pages}{e2208975119}.
\newblock


\bibitem[Yuan and Altenburger(2022)]%
        {yuan2022two}
\bibfield{author}{\bibinfo{person}{Yuan Yuan} {and} \bibinfo{person}{Kristen~M
  Altenburger}.} \bibinfo{year}{2022}\natexlab{}.
\newblock \bibinfo{title}{A Two-Part Machine Learning Approach to
  Characterizing Network Interference in A/B Testing}.
\newblock
\newblock


\end{thebibliography}

\appendix

\section{Proofs}

\subsection{Proof of Proposition \ref{prop: prop1}}
\begin{proof}
Let's rewrite the estimator $\widehat{\tau}^{1}$ as
\begin{align*}
\widehat{\tau}^{1} = & \frac{1}{n} \sum_{i=1}^n\frac{W_iY_i}{\pi} - \frac{1}{n} \sum_{i=1}^n \frac{(1-W_i)Y_i}{1-\pi} \\
= & \frac{1}{n\pi} \sum_{i=1}^n W_i \left[\sum_{j=1}^n \Big( \alpha_{j,i} + \beta_{j,i} W_j + \gamma_{j,i} W_i + \zeta_{j,i} W_i W_j \Big) \right] - \\
& \,\, \frac{1}{n(1-\pi)} \sum_{i=1}^n (1-W_i) \left[\sum_{j=1}^n \Big( \alpha_{j,i} + \beta_{j,i} W_j + \gamma_{j,i} W_i + \zeta_{j,i} W_i W_j \Big) \right] \\
=& \frac{1}{n\pi} \sum_{i \neq j} W_i \Big( \alpha_{j,i} + \beta_{j,i} W_j + \gamma_{j,i} + \zeta_{j,i} W_j\Big) - \\
& \,\, \frac{1}{n(1-\pi)} \sum_{i \neq j} (1-W_i) \Big( \alpha_{j,i} + \beta_{j,i} W_j \Big)
\end{align*}
Note that
\begin{enumerate}
    \item $\mathbb{E}(W_i) = \pi$.
    \item $\{W_i\}_{i=1}^n$ are independent, $\mathbb{E}(W_iW_j) = \pi^2$ for $i \neq j$. 
\end{enumerate}
So we have
\begin{align*}
\mathbb{E}(\widehat{\tau}^{1}) = & \frac{1}{n} \sum_{i \neq j} \Big( \alpha_{j,i} + \beta_{j,i} \pi + \gamma_{j,i} + \zeta_{j,i}\pi\Big) - \frac{1}{n} \sum_{i \neq j} \Big( \alpha_{j,i} + \beta_{j,i} \pi \Big) \\
= & \frac{1}{n} \sum_{i \neq j} \Big( \gamma_{i,j} + \zeta_{i,j}\pi \Big).
\end{align*}
\end{proof}

\subsection{Proof of Proposition \ref{prop: prop2}}
\begin{proof}
Follow the same steps,
\begin{align*}
\widehat{\tau}^{2} = &  \frac{1}{n} \sum_{i=1}^n\frac{W_iD_i}{\pi} - \frac{1}{n} \sum_{i=1}^n \frac{(1-W_i)D_i}{1-\pi} \\
= &  \frac{1}{n\pi} \sum_{i=1}^n W_i \left[\sum_{j=1}^n \Big( \alpha_{i,j} + \beta_{i, j} W_i + \gamma_{i, j} W_j + \zeta_{i,j} W_i W_j\Big) \right] - \\
& \,\, \frac{1}{n(1-\pi)} \sum_{i=1}^n (1-W_i) \left[\sum_{j=1}^n \Big( \alpha_{i,j} + \beta_{i, j} W_i + \gamma_{i, j} W_j + \zeta_{i,j} W_i W_j \Big) \right] \\
=&  \frac{1}{n\pi} \sum_{i \neq j}^n W_i  \Big( \alpha_{i,j} + \beta_{i, j} + \gamma_{i, j} W_j + \zeta_{i,j} W_j \Big)  - \\
& \,\, \frac{1}{n(1-\pi)} \sum_{i \neq j} (1-W_i)  \Big(\alpha_{i,j} + \gamma_{i, j} W_j \Big). \\
\end{align*} 
Again, $\{W_i\}_{i=1}^n$ are independent. Hence
\begin{align*}
\mathbb{E} (\widehat{\tau}^{2}) = & \frac{1}{n} \sum_{i \neq j}\Big( \alpha_{i,j} + \beta_{i, j} + \gamma_{i, j} \pi + \zeta_{i,j} \pi \Big) - \frac{1}{n} \sum_{i \neq j} \Big(\alpha_{i,j} + \gamma_{i, j} \pi \Big) \\
= & \frac{1}{n} \sum_{i \neq j} \Big(\beta_{i, j} + \zeta_{i,j} \pi\Big).
\end{align*} 
\end{proof}

\subsection{Proof of Proposition \ref{prop: prop4}}
\begin{proof}
Follow the same steps,
\begin{align*}
\widehat{\tau}^{1} = &  \frac{1}{n} \sum_{i=1}^n \frac{ V_i W_i Y_i }{ p\pi } - \frac{1}{n} \sum_{i=1}^n \frac{ V_i (1-W_i) Y_i }{ p(1-\pi) } \\
= &  \frac{1}{np\pi} \sum_{i=1}^n V_i W_i \left[\sum_{j=1}^n \Big(\alpha_{j,i} + \beta_{j,i} W_j + \gamma_{j,i} W_i + \zeta_{j,i} W_i W_j  \Big)\right] - \\
& \,\, \frac{1}{np(1-\pi)} \sum_{i=1}^n V_i(1-W_i) \left[\sum_{j=1}^n \Big(\alpha_{j,i} + \beta_{j,i} W_j + \gamma_{j,i} W_i + \zeta_{j,i} W_i W_j  \Big)\right] \\
=&  \frac{1}{np\pi} \sum_{i \neq j}^n V_i W_i  \Big( \alpha_{j,i} + \beta_{j,i} W_j + \gamma_{j,i}+ \zeta_{j,i} W_j  \Big) - \\
& \,\, \frac{1}{np(1-\pi)} \sum_{i \neq j} V_i(1-W_i)  \Big( \alpha_{j,i} + \beta_{j,i} W_j \Big) \\
\end{align*} 
Note that
\begin{enumerate}
    \item $\mathbb{E}(V_iW_i) = p\pi$.
    \item $\mathbb{E}(V_iW_iW_j) = p^2\pi$.
    \item $\mathbb{E}\big(V_i(1-W_i)\big) = p(1-\pi)$.
    \item $\mathbb{E}\big(V_i(1-W_i)W_j\big) = p^2\pi(1-\pi)$.
\end{enumerate}
Hence
\begin{align*}
\mathbb{E}(\widehat{\tau}^{1}) = & \frac{1}{n} \sum_{i \neq j} \Big( \alpha_{j,i} + \beta_{j,i} p\pi + \gamma_{j,i}+ \zeta_{j,i} p\pi  \Big) - \frac{1}{n} \sum_{i \neq j} \Big( \alpha_{j,i} + \beta_{j,i} p\pi \Big)  \\
 = & \frac{1}{n} \sum_{i\neq j}\Big(\gamma_{j,i} + p\pi\zeta_{j,i} \Big).
\end{align*} 
\end{proof}

\subsection{Proof of Proposition \ref{prop: prop5}}
\begin{proof}
Follow the same steps,
\begin{align*}
\widehat{\tau}^{2} = & \frac{1}{n} \sum_{i=1}^n\frac{ V_i W_i D_i }{p\pi} - \frac{1}{n} \sum_{i=1}^n \frac{ V_i (1-W_i) D_i }{p(1-\pi)} \\
= &  \frac{1}{np\pi} \sum_{i=1}^n V_iW_i \left[\sum_{j=1}^n \Big(\alpha_{i,j} + \beta_{i, j} W_i + \gamma_{i, j} W_j + \zeta_{i,j} W_i W_j\Big) \right] - \\
& \,\, \frac{1}{np(1-\pi)} \sum_{i=1}^n V_i(1-W_i) \left[\sum_{j=1}^n \Big(\alpha_{i,j} + \beta_{i, j} W_i + \gamma_{i, j} W_j + \zeta_{i,j} W_i W_j\Big) \right] \\
=&  \frac{1}{np\pi} \sum_{i \neq j} V_iW_i  \Big(\alpha_{i,j} + \beta_{i, j}+ \gamma_{i, j} W_j + \zeta_{i,j} W_j\Big) - \\
& \,\ \frac{1}{np(1-\pi)} \sum_{i \neq j} V_i(1-W_j)  \Big(\alpha_{i,j} + \gamma_{i, j} W_j\Big) \\
\end{align*} 
Hence
\begin{align*}
\mathbb{E}(\widehat{\tau}^{2}) = &  \frac{1}{n} \sum_{i \neq j} \Big(\alpha_{i,j} + \beta_{i, j}+ \gamma_{i, j} p\pi + \zeta_{i,j} p\pi\Big) - \frac{1}{n} \sum_{i\neq j} \Big(\alpha_{i,j} + \gamma_{i, j} p\pi \Big) \\
= & \frac{1}{n} \sum_{i\neq j}\Big(\beta_{i,j}+ p\pi\zeta_{i,j}\Big)
\end{align*} 
\end{proof}

\subsection{Proof of Proposition \ref{prop: prop6}}
\begin{proof}
Follow the same steps,
\begin{align*}
\widehat{\tau}^{1} = & \frac{1}{n} \sum_{i=1}^n \frac{ V_i W_i Y_i }{ p\pi } - \frac{1}{n} \sum_{i=1}^n \frac{ V_i (1-W_i) Y_i }{ p(1-\pi) } \\
= &  \frac{1}{np\pi} \sum_{i=1}^n V_i W_i \left[\sum_{j=1}^n \Big(\alpha_{j,i} + \beta_{j,i} W_j + \gamma_{j,i} W_i + \zeta_{j,i} W_i W_j  \Big)\right] - \\
& \,\, \frac{1}{np(1-\pi)} \sum_{i=1}^n V_i(1-W_i) \left[\sum_{j=1}^n \Big(\alpha_{j,i} + \beta_{j,i} W_j + \gamma_{j,i} W_i + \zeta_{j,i} W_i W_j  \Big)\right] \\
=&  \frac{1}{np\pi} \sum_{i \neq j} V_i W_i \Big( \alpha_{j,i} + \beta_{j,i} W_j + \gamma_{j,i}+ \zeta_{j,i} W_j  \Big) - \\
& \,\, \frac{1}{np(1-\pi)} \sum_{i \neq j} V_i(1-W_i) \Big( \alpha_{j,i} + \beta_{j,i} W_j \Big) \\
\end{align*}
Note that
\begin{enumerate}
    \item $\mathbb{E}(V_iW_i) = p\pi, \mathbb{E}\big(V_i(1-W_i)\big)=p(1-\pi).$
    \item $\mathbb{E}\big(V_iW_iW_j\big) = \mathbb{E}[ \mathbb{E}\big(V_iW_iW_j\big | j\text{ and }  i  \text{ are in the same cluster}) ] = \bar{\sigma} * \mathbb{E}\big(V_iW_i\big) + (1-\bar{\sigma}) * \mathbb{E}\big(V_iW_i\big)\mathbb{E}\big(W_j\big) = \bar{\sigma}p\pi + (1-\bar{\sigma})p^2\pi^2.$
    \item $\mathbb{E}\big(V_i(1-W_i)W_j\big) = \mathbb{E}[ \mathbb{E}\big(V_i(1-W_i)W_j\big | j \text{ and } i $  are not in the same cluster $) ] = (1-\bar{\sigma})p^2\pi(1-\pi)$.
\end{enumerate}

Hence
\begin{align*}
\mathbb{E}(\widehat{\tau}^{1}) = & \frac{1}{n} \sum_{i \neq j} \Big( \alpha_{j,i} + \gamma_{j,i} + (\beta_{j,i} + \zeta_{j,i}) ( \bar{\sigma} +(1-\bar{\sigma})p\pi )  \Big) - \\
& \,\, \frac{1}{n} \sum_{i \neq j} \Big( \alpha_{j,i} + \beta_{j,i} (1-\bar{\sigma})p\pi \Big) \\
 = & \frac{1}{n} \sum_{i \neq j} \Big( \beta_{i,j}\bar{\sigma} + \gamma_{i,j} + \big(\bar{\sigma}+(1-\bar{\sigma})p\pi \big)\zeta_{i,j} \Big)
\end{align*} 
\end{proof}

\subsection{Proof of Proposition \ref{prop: prop7}}
\begin{proof}
Follow the same steps in (A.5),
\begin{align*}
\widehat{\tau}^{2} = & \frac{1}{n} \sum_{i=1}^n \frac{ V_i W_i D_i }{ p\pi } - \frac{1}{n} \sum_{i=1}^n \frac{ V_i (1-W_i) D_i }{ p(1-\pi) } \\
= &  \frac{1}{np\pi} \sum_{i=1}^n V_i W_i \left[\sum_{j=1}^n \Big(\alpha_{i,j} + \beta_{i,j} W_i + \gamma_{i,j} W_j + \zeta_{i,j} W_i W_j  \Big)\right] - \\
& \,\, \frac{1}{np(1-\pi)} \sum_{i=1}^n V_i(1-W_i) \left[\sum_{j=1}^n \Big(\alpha_{i,j} + \beta_{i,j} W_i + \gamma_{i,j} W_j + \zeta_{i,j} W_i W_j  \Big)\right] \\
=&  \frac{1}{np\pi} \sum_{i \neq j} V_i W_i \Big( \alpha_{i,j} + \beta_{i,j} + \gamma_{i,j}W_j+ \zeta_{i,j} W_j  \Big) - \\
& \,\, \frac{1}{np(1-\pi)} \sum_{i \neq j} V_i(1-W_i) \Big( \alpha_{i,j} + \gamma_{i,j} W_j \Big) \\
\end{align*}

Hence
\begin{align*}
\mathbb{E}(\widehat{\tau}^{2}) = & \frac{1}{n} \sum_{i \neq j} \Big( \alpha_{i,j} + \beta_{j,i} + (\gamma_{j,i} + \zeta_{i,j}) ( \bar{\sigma} +(1-\bar{\sigma})p\pi )  \Big) - \\
& \,\, \frac{1}{n} \sum_{i \neq j} \Big( \alpha_{i,j} + \gamma_{i,j} (1-\bar{\sigma})p\pi \Big) \\
 = & \frac{1}{n} \sum_{i \neq j} \Big( \beta_{i,j} + \gamma_{i,j}\bar{\sigma} + \big(\bar{\sigma}+(1-\bar{\sigma})p\pi \big)\zeta_{i,j} \Big)
\end{align*} 
\end{proof}

\end{document}